\documentclass[11pt]{article}
\usepackage{amsmath,amssymb,mathrsfs,url}
\usepackage{cite}

\title{Loop quantum cosmology and singularities}

\author{Ward Struyve\footnote{Mathematisches Institut, Ludwig-Maximilians-Universit\"at M\"unchen, Theresienstr.\ 39, 80333 M\"unchen, Germany. E-mail: ward.struyve@gmail.com}
}

\newcommand{\dd}{\mbox{d}}

\def\lam{\lambda}

\def\pa{\partial}

\def\th{\theta}
\def\al{\alpha}

\def\ka{\kappa}
\def\ii{\textrm i}
\def\ee{\textrm e}

\newcommand{\be}{\begin{equation}}
\newcommand{\en}{\end{equation}}

\begin{document}
\maketitle

\begin{abstract}
\noindent
Loop quantum gravity is believed to eliminate singularities such as the big bang and big crunch singularity. This belief is based on studies of so-called loop quantum cosmology which concerns symmetry-reduced models of quantum gravity. In this paper, the problem of singularities is analysed in the context of the Bohmian formulation of loop quantum cosmology.  In this formulation there is an actual metric in addition to the wave function, which evolves stochastically (rather than deterministically as the case of the particle evolution in non-relativistic Bohmian mechanics). Thus a singularity occurs whenever this actual metric is singular. It is shown that in the loop quantum cosmology for a homogeneous and isotropic Friedmann-Lema\^itre-Robertson-Walker space-time with arbitrary constant spatial curvature and cosmological constant, coupled to a massless homogeneous scalar field, a big bang or big crunch singularity is never obtained. This should be contrasted with the fact that in the Bohmian formulation of the Wheeler-DeWitt theory singularities may exist.
\end{abstract}

\section{Introduction}
According to general relativity space-time singularities such as a big bang or big crunch are generic. This is often taken as signaling a limit on the validity of the classical theory of gravity. The hope is that a quantum theory for gravity will eliminate the singularities. Several candidates for a quantum gravity theory have been proposed, such as the Wheeler-DeWitt theory, loop quantum gravity, string theory, etc.\ \cite{kiefer04}. These different proposals may lead to different answers to the question of singularities. Much effort has gone into studying mini-superspace models, which are symmetry-reduced versions of quantum gravity, and which are obtained by using the usual quantization techniques on symmetry-reduced general relativity. In particular, in recent years, there has been a comparison of the Wheeler-DeWitt theory and loop quantum gravity (called loop quantum cosmology (LQC) in this context) in the case of a homogeneous and isotropic Friedmann-Lema\^itre-Robertson-Walker (FLRW) metric coupled to a scalar field. It was found that (for a large class of wave functions) the Wheeler-DeWitt theory yields singularities while LQC has no singularities \cite{ashtekar06b,ashtekar06c,ashtekar06d,ashtekar08}. However, there are some problems which have to do with applying standard quantum theory in this case. First of all there is the measurement problem, which has to do with the ambiguity of when exactly collapses happen. This problem carries over from non-relativistic quantum mechanics and is especially severe in the context of quantum cosmology. Namely, the aim is to describe the whole universe (albeit with simplified models) and then there are no outside observers or measurement devices that could collapse the wave function. In addition, the aim is also to describe for example the early universe and there are no observers or measurement devices present even within the universe. Second, there is the problem of time \cite{isham92,kuchar92,kiefer04}. In both the Wheeler-DeWitt theory and LQC, the wave function is static. So how can time evolution can be explained in terms of such a wave function? How can we tell from the theory whether the universe is expanding or contracting or running into a singularity? Finally, there is the problem of what it means to have a space-time singularity. In both theories, the universe is described solely by a wave function, but there is no actual metric. Various definitions of what a singularity could mean have been explored \cite{dewitt67a,ashtekar06b,ashtekar06c,ashtekar06d,ashtekar08,ashtekar11}: that the wave function has support on singular metrics, that the wave function is peaked around singular metrics, that the expectation value of the metric operator is singular, etc. Although these definitions may have something so say about the occurrence or non-occurrence singularities, neither of these is completely satisfactory. In fact, since there is merely the wave function, one might even consider the question about space-time singularities as off-target, since it is the dynamics of the wave function that needs to be well-defined.

Various possible solutions have been explored to solve (some) of these problems. In particular, a number of solutions to the measurement problem exist, such as for example the Many Worlds theory, spontaneous collapse models and Bohmian mechanics. There also exist a number of approaches to solving the problem of time, for a recent overview see \cite{anderson12}. Solving one problem may also lead to the solution of another one. For example, in spontaneous collapse models the collapses are objective processes. But the collapses entail change and hence may solve the problem of time. The question of singularities in the context of both the Wheeler-DeWitt theory and LQC has been discussed in great detail for the Consistent Histories approach to quantum mechanics \cite{craig10,craig11,falciano15,craig16}.

In this paper, we consider Bohmian mechanics. Bohmian mechanics is an alternative to standard quantum mechanics that solves the aforementioned problems. In non-relativistic Bohmian mechanics there are particles in addition to the wave function \cite{bohm93,holland93b,duerr09}. The wave function determines the motion of the particles in a way that is similar to the way the Hamiltonian determines the motion of classical particles. There is no measurement problem since there is no collapse of the wave function. Outcomes of measurements are determined by the positions of the actual particles. We explore the question of singularities in the mini-superspace model of a FLRW space-time coupled to a homogeneous scalar field in the context of Bohmian mechanics. In the Bohmian versions of the Wheeler-DeWitt theory there is an actual FLRW metric and scalar field whose dynamics is determined by the wave function in a deterministic way \cite{shtanov96,goldstein04,pinto-neto05,pinto-neto13}. In the Bohmian version of LQC, which is developed here, there are also an actual FLRW metric and scalar field, but now the dynamics of the metric is stochastic rather than deterministic. While the wave function is static, the actual metric and scalar field generically evolve in time. The wave function does not collapse, although it may at an effective level, so that there is no measurement problem. Finally, it is also clear in this case what is meant by a singularity: there are singularities whenever the actual metric becomes singular.

In previous work \cite{pinto-neto12b,falciano15}, the question of singularities was studied for the Bohmian version of the Wheeler-DeWitt theory for mini-superspace. It was found that there may or may not be singularities; it depends on the wave function and the initial conditions of the actual fields. In particular, there are wave functions for which there are no singularities for any of the initial conditions and there wave functions for which there are always singularities. In this paper, we develop a Bohmian theory for LQC and consider the question of singularities. We consider some common models for LQC which correspond to different wave equations (arising from operator ordering ambiguities) and find that big bang or big crunch singularities do not occur for any value of the spatial curvature and cosmological constant. 

The outline of the paper is as follows. First, we consider the Wheeler-DeWitt quantization of the mini-superspace model, the corresponding Bohmian theory and the results for singularities. Then in section \ref{lqg}, we present some common models for LQC. In section \ref{bohmian lqg}, we present their Bohmian versions and show that there is no big bang or big crunch singularity. In section \ref{time}, we discuss how the problem of time is usually addressed in LQC and compare it to the Bohmian solution. Finally, in section \ref{modified}, we consider a modified Wheeler-DeWitt equation inspired by loop quantum cosmology which also has the potential to eliminate singularities.

\section{Wheeler-DeWitt quantization}\label{wdwbm}
A classical FLRW space-time is described by a metric
\be
\dd s^2 = N(t)^2 \dd t^2 - a(t)^2  \dd \Omega^2_k ,
\label{1}
\en
where $N>0$ is the lapse function, $a=\ee^{\al}$ is the scale factor, and $\dd \Omega^2_k$ is the spatial line-element on three-space with constant curvature $k$. The coupling to a homogeneous scalar field $\phi$ is described by the Lagrangian
\be
L = N \ee^{3\al} \left( \ka^2 \frac{\dot \phi^2}{2N^2} - \ka^2 V_M  - \frac{\dot \al^2}{2N^2} - V_G\right),
\label{1.1}
\en
where $\kappa = \sqrt{4\pi G/3}$, with $G$ the gravitational constant, $V_M$ is the potential for the scalar field, $V_G =- \frac{1}{2}k\ee^{-2\al} + \frac{1}{6}\Lambda$, and $\Lambda$ is the cosmological constant \cite{halliwell91b,struyve15}. The classical equations of motion are
\be
\frac{d}{dt}\left( \frac{\ee^{3\al} \dot \phi}{N}\right) + N \ee^{3\al} \pa_\phi V_M =0,
\label{1.2}
\en
\be
 \frac{\dot \al^2}{N^2} = 2 \ka^2 \left( \frac{\dot \phi^2}{2N^2} + V_M\right) + 2 V_G.
 \label{1.3}
 \en
The latter equation is the Friedmann equation. The Friedmann acceleration equation follows from \eqref{1.2} and \eqref{1.3}. $N$ remains an arbitrary function of time. This implies that the dynamics is time reparameterization invariant. 

Canonical quantization of the classical theory leads to the Wheeler-DeWitt equation
\be
\left[- \frac{1}{2\ee^{3\al}} \pa^2_\phi + \frac{\ka^2}{2\ee^{3\al}}\pa^2_\al + \ee^{3\al}\left(V_M + \frac{1}{\ka^2} V_G\right) \right]\psi(\phi,\al) =0.
\label{1.4}
\en
In the context of standard quantum theory, this equation is hard to interpret due to the problem of time \cite{isham92,kuchar92,kiefer04}. Namely, the wave function is static. So how can the apparent time evolution of the universe be accounted for?

In the Bohmian theory \cite{vink92,pinto-neto12b}, there is an actual scalar field $\phi$ and an actual FLRW metric of the form \eqref{1}, whose time evolution is determined by
\be
\dot \phi = \frac{N}{e^{3\alpha}} \pa_\phi S , \quad \dot \al = - \frac{N}{e^{3\alpha}} \kappa^2 \pa_\al S ,
\label{1.5}
\en
where $\psi = |\psi| \ee^{\ii S}$. The function $N$ is again the lapse function, which is arbitrary, and, just as in the classical case, implies that the dynamics is time reparameterization invariant. 

As usual, the Bohmian dynamics can be motivated by the conservation equation
\be
\pa_\phi J_\phi + \pa_\al J_\al = 0 ,
\en
where 
\be
J_\phi = \pa_\phi S |\psi|^2, \qquad J_\al = -\pa_\al S |\psi|^2.
\en
We then take $(\dot \phi, \dot \al) \sim (J_\phi,J_\al)$. The natural proportionality constant is given by $N/e^{3\alpha}|\psi|^2$, since it follows from the equations of motion \eqref{1.5} that 
\be
\frac{d}{dt}\left( \frac{\ee^{3\al} \dot \phi}{N}\right) + N \ee^{3\al} \pa_\phi (V_M + Q_M)=0,
\en
\be
\frac{\dot \al^2}{N^2} = 2 \ka^2 \left( \frac{\dot \phi^2}{2N^2} +(V_M + Q_M) \right) + 2 (V_G + Q_G),
\en
where
\be
Q_M = -\frac{1}{2\ee^{6\al}} \frac{\pa^2_\phi |\psi|}{|\psi|} , \qquad Q_G = \frac{\kappa^4}{2\ee^{6\al}} \frac{\pa^2_\al |\psi|}{|\psi|}
\en
are respectively the matter and the gravitational quantum potential. As such, the classical equations are obtained, with addition of the quantum potentials to the classical potentials. The guidance equations can also be obtained from the classical Hamilton equations by replacing the conjugate momenta $\pi_\al$ and $\pi_\phi$ by respectively $\pa_\al S$ and $\pa_\phi S$ (a procedure that works for Hamiltonians that are at most quadratic in the momenta \cite{struyve10}). 

Even though the wave function is static, the Bohmian scale factor generically depends on time. As such there is no problem of time. We will discuss this further in section \ref{time}.

The so-called quantum equilibrium measure is $\ee^{3\al}|\psi(\phi,\al)|^2 d\phi d\al$ (or $a^2|\psi(\phi,a)|^2da d\phi$). This measure is preserved by the Bohmian dynamics. However, it is non-normalizable (i.e., no probability measure), so it can not straightforwardly be used to extract probabilities for possible histories (while in non-relativistic Bohmian mechanics the equilibrium measure gives rise to Born's law). Probabilities are only secondary, with the primary role of the wave function to determine the evolution of the metric and the scalar field. For that reason, it is also not important to introduce a Hilbert space. We just need to assume that the wave function is such that the Bohmian dynamics is well-defined. 

Since it is rather unclear what the Wheeler-DeWitt equation means in the context of standard quantum mechanics, there is also no straightforward comparison between the Bohmian predictions and those of standard quantum theory possible.  This is unlike the situation in non-relativistic quantum mechanics, where it can be shown that Bohmian mechanics reproduces the predictions of standard quantum theory (provided the latter are unambiguous). For example, the Hartle-Hawking wave function which is studied in great detail is empirically inadequate from the Bohmian point of view since it is a real wave function and implies a stationary universe \cite{shtanov96}.

Let us now turn to the question of singularities. In the classical theory, there is a big bang or big crunch singularity when $a=0$. This singularity is obtained for generic solutions. For example, in the case of $V_M = V_G = 0$, the classical equations lead to
\be
\dot \phi =  \frac{N}{\ee^{ 3\al}} c , \qquad \dot \al = \pm  \frac{N}{\ee^{ 3\al}} \kappa^2 c ,
\label{2}
\en
where $c$ is an integration constant. In the case $c=0$, the universe is static and described by the Minkowski metric. In this case there is no singularity. For $c \neq 0$, we have
\be
\al = \pm \kappa^2 \phi + {\bar c},
\label{3}
\en
with ${\bar c}$ another integration constant. In terms of proper time $\tau$ for a co-moving observer (i.e.\ moving with the expansion of the universe), which is defined by $d\tau = N dt$, integration of \eqref{2} yields $a=e^\al=\left[3(c\tau + {\tilde c}) \right]^{1/3}$, where ${\tilde c}$ is an integration constant, so that $a=0$ for $\tau = - {\tilde c}/c$ (and there is a big bang if $c>0$ and a big crunch if $c<0$). This means that the universe reaches the singularity in finite proper time.

In the standard quantum mechanical approach to the Wheeler-DeWitt theory, the complete description is given by the wave function and as such, as mentioned in the introduction, the notion of a singularity becomes ambiguous. Not so in the Bohmian theory. The Bohmian theory describes the evolution of an actual metric and hence there are singularities whenever this metric is singular. The question of singularities in the special case where $V_M = V_G = 0$ was considered in \cite{pinto-neto12b,falciano15}. In this case, the Wheeler-DeWitt equation is
\be
\frac{1}{a^3}\pa^2_\phi \psi - \kappa^2 \frac{1}{a^2} \pa_a (a \pa_a \psi) = 0, 
\label{3.1}
\en
or in terms of $\al$:
\be
\pa^2_\phi \psi - \kappa^2 \pa^2_\al \psi = 0 .
\label{4}
\en
The solutions are
\be
\psi = \psi_R(\ka \phi - \al) + \psi_L(\ka \phi + \al).
\label{5}
\en
The actual metric might be singular; it depends on the wave function and on the initial conditions. For example, for a real wave function, $S=0$, and hence the universe is static, so that there is no singularity. Assuming that $\lim_{x \to \pm \infty} |\psi_{R,L}(x)| = 0$, it can be shown (work in preparation with D.\ D\"urr and H.\ Ochner) that the only wave functions for which there are no singularities are of the form 
\be
\psi(\phi,\al) = |\psi_R(\ka \phi - \al)| + |\psi_L(\ka \phi + \al)|\ee^{\ii \th}
\label{6}
\en
(up to an irrelevant constant phase factor) with $\th$ a constant. On the other hand, for wave functions $\psi=\psi_{R,L}$ the solutions are always classical, i.e., they are either static (if $\pa_\al S(\ka \phi(0) - \al(0)) =0$, with $(\phi(0),\al(0))$ the initial configuration) or they reach a singularity in finite proper time $\tau$ for a co-moving observer (if $\pa_\al S(\ka \phi(0) - \al(0)) \neq 0)$. Wave functions with $\psi_R=\psi_L$ satisfy $\psi(\phi,\al) = -\psi(\phi,-\al)$ and lead to trajectories that do not cross the plane $\al =0$ in $(\phi,\al)$-space. As such trajectories starting with $\al(0) > 0$ will not have singularities.  

In comparison we note that in the context of the consistent histories approach to quantum mechanics, it was shown that singularities are always obtained for this system \cite{craig10,craig11,falciano15,craig16}.

Finally, in order to compare to LQC, we introduce the variable $\nu = \epsilon C a^3$, with $\epsilon=\pm1$ and $C>0$ a constant given in \eqref{10}. The Wheeler-DeWitt equation then reads
\be
\frac{1}{|\nu|}\pa^2_\phi \psi - 9\kappa^2 \pa_\nu (|\nu| \pa_\nu \psi) = 0 
\label{7}
\en
and the guidance equations read
\be
\dot \phi = \frac{NC}{|\nu|} \pa_\phi S , \quad \dot \nu = - N9C \kappa^2 |\nu| \pa_\nu S .
\label{8}
\en
The quantum equilibrium measure is $|\psi(\phi,\nu)|^2 d\phi d\nu$. In analogy with LQC we can further assume $\psi(\phi,-\nu)=\psi(\phi,\nu)$. (While $\psi(\phi,-\nu)=\psi(\phi,\nu)$ actually introduces the boundary condition $\pa_\nu \psi(\phi,0) = 0$, this is not important since we will be making the comparison only for large $|\nu|$.)

\section{Loop quantum cosmology}\label{lqg}
Loop quantization is a different way to quantize general relativity \cite{rovelli04,rovelli14}. Application of  this quantization method to the mini-superspace model considered here results in the following. States are functions $\psi_\nu(\phi)$ of a continuous variable $\phi$ and a discrete variable 
\be
\nu = \epsilon C a^3, 
\label{9}
\en
with
\be
C = \frac{ V_0}{2\pi G \gamma },
\label{10}
\en
where $\epsilon = \pm 1$ is the orientation of the triad (which is used in passing from the metric representation of general relativity to the connection representation), $V_0$ is the fiducial volume (which is introduced to make volume integrations finite) and $\gamma$ is the Barbero-Immirzi parameter. $\nu$ is discrete as it is given by $\nu = 4 n\lambda$ with $n \in {\mathbb Z}$ and $\lambda^2 = 2 \sqrt{3}\pi\gamma G$. The value $\nu=0$, which corresponds to the singularity, is included. One could also take $\nu = \epsilon + 4 n\lambda$, with $\epsilon \in (0,4\lam)$. This does not include the value $\nu=0$ and as such the singularity would automatically be avoided in the corresponding Bohmian theory (because, as will be discussed in the next section, in the Bohmian theory the possible values the scale factor can take are given by the discrete values of $\nu$ on which $\psi$ has its support).

As usual, the quantization is not unique. Because of operator ordering ambiguities different wave equations may be obtained. We discuss 4 different ones that are commonly used in the literature: the Ashtekar-Pawlowski-Singh (APS) model \cite{ashtekar06d,ashtekar08}, the simplified APS model \cite{ashtekar08}, called sLQC, the Mart\'in-Benito--Mena-Marug\'an--Olmedo (MMO) model \cite{martin-benito09} and the simplified MMO model \cite{martin-benito09}, called sMMO \cite{mena-marugan11,banerjee12}. A comparison of these models can be found in \cite{mena-marugan11}. (The APS model is an improved version of an earlier model of Ashtekar, Pawlowski and Singh \cite{ashtekar06b,ashtekar06c}. Unlike the earlier version, the APS model tends to yield a bounce when the matter density enters the Planck regime rather than higher energies. The Bohmian singularity analysis of this model would not differ much from that of the APS model. In the limit for large $\nu$ this model does not give rise to the Wheeler-DeWitt equation \eqref{7} but to one that differs from it by a factor ordering. The models for LQC that we consider here will all give rise to the Wheeler-DeWitt equation \eqref{7}.)

In all models, the wave equation is of the form
\be 
B_\nu \pa^2_\phi \psi_\nu(\phi) + \sum_{\nu'} K_{\nu,\nu'} \psi_{\nu'}(\phi) = 0 ,
\label{11}
\en
with $\psi_\nu =\psi_{-\nu} $ and $B_\nu$ and $K_{\nu,\nu'}=K_{\nu',\nu}$ are real. The gravitational part, determined by $K$, is not a differential equation but a difference equation. For now, we do not consider a non-zero curvature or a cosmological constant. This will be done at the end of this section. Just as in the case of the Wheeler-DeWitt theory, we will not worry about a suitable Hilbert space for the wave equation.  

In the APS model, the wave equation is 
\be
B_\nu \pa^2_\phi \psi_\nu(\phi)  - 9 \kappa^2 D_{2\lam} (|\nu| D_{2\lam} \psi_\nu(\phi)) = 0,
\label{12}
\en
where
\be
D_h \psi_\nu = \frac{\psi_{\nu + h} - \psi_{\nu - h}}{2h},
\label{13}
\en
so that  
\be
K_{\nu,\nu\pm 4\lam} = - \frac{9 \kappa^2}{16\lam^2} |\nu \pm 2 \lam| \,, \qquad   \qquad  K_{\nu,\nu} = - K_{\nu,\nu+4\lam} -   K_{\nu,\nu-4\lam}      
\label{14}
\en
and the other $K_{\nu,\nu'}$ are zero. Various choices for $B_\nu$ exist, again due to operator ordering ambiguities \cite{bojowald02,bojowald08}. One choice is \cite{ashtekar06d}:
\be
B_\nu =  |\nu| \left|3 D_\lam|\nu |^{1/3} \right|^3 = |\nu|  \left|3 \frac{|\nu + \lam|^{1/3} - |\nu - \lam|^{1/3} }{2\lam}\right|^3 .
\label{15}
\en
Another one is \cite{ashtekar08}:
\be
B_\nu =  \left|\frac{3}{2}D_\lam|\nu |^{2/3} \right|^3 = \left|\frac{3}{2} \frac{|\nu + \lam|^{2/3} - |\nu - \lam|^{2/3} }{2\lam}\right|^3.
\label{16}
\en
All choices of $B_\nu$ share the important properties that $B_0=0$ and that for $|\nu| \gg \lam$ (taking the limit $\lam \to 0$, or equivalently, taking the Barbero-Immirzi parameter or the area gap to zero), $B_\nu \to 1/|\nu|$. 

In sLQC, the wave equation takes the form \eqref{12}, but with
\be
B_\nu = \frac{1}{|\nu|} .
\label{17}
\en
In this case, $B_0 \neq 0$. This is a simplification that was introduced to make the model exactly solvable.

In the MMO model, the wave equation is 
\be
B_\nu \pa^2_\phi \psi_\nu(\phi)  - 9 \kappa^2 \sqrt{B_\nu} \left(G_\nu D_{2\lam} G_\nu   \right)^2 \sqrt{B_\nu}\psi_\nu(\phi) =0,
\label{18}
\en
where
\begin{equation}
G_\nu = \left\{ 
\begin{array}{ll}
|\nu|^{1/3} B_\nu^{-1/6}  & \quad \text{if } \nu \neq 0\\
0 &\quad  \text{if } \nu = 0\
\end{array} \right.,
\label{19}
\end{equation}
with $B_\nu$ given by one of the choices for the APS model. Hence 
\be
K_{\nu,\nu \pm  4\lam} =  - \frac{9 \kappa^2}{16\lam^2} \sqrt{B_\nu}G_\nu G^2_{\nu \pm 2\lam} G_{\nu \pm 4\lam}\sqrt{B_{\nu \pm 4\lam}},
\label{20}
\en
\be
K_{\nu,\nu} = \frac{9 \kappa^2}{16\lam^2} \sqrt{B_\nu}G_\nu \left( G^2_{\nu + 2\lam}G_{\nu - 2\lam}\sqrt{B_{\nu -2\lam}}  + G^2_{\nu - 2\lam}G_{\nu + 2\lam}\sqrt{B_{\nu + 2\lam}} \right).
\label{21}
\en
This model has the special property that $K_{\nu, 0} = K_{0, \nu} = 0$. This implies that $\psi_0(\phi)$ is dynamically decoupled from the $\psi_\nu(\phi)$ with $\nu \neq 0$ and moreover that $\psi_0(\phi)$ is arbitrary. 

The sMMO model is obtained from the MMO model by replacing $G_\nu$ by $\sqrt{\nu}$ in \eqref{18} (which amounts to replacing $B_\nu$ by $1/|\nu|$ in \eqref{19}). This results in the wave equation 
\be
B_\nu \pa^2_\phi \psi_\nu(\phi)  - 9 \kappa^2 \sqrt{B_\nu} \left(\sqrt{|\nu|} D_{2\lam}  \sqrt{|\nu|}  \right)^2 \sqrt{B_\nu}\psi_\nu(\phi) =0,
\label{22}
\en
so that 
\be
K_{\nu,\nu \pm  4\lam} =  - \frac{9 \kappa^2}{16\lam^2} \sqrt{|\nu| B_\nu}  |\nu \pm 2\lam| \sqrt{|\nu\pm 4\lam |B_{\nu \pm 4\lam}},
\label{23}
\en
\be
K_{\nu,\nu} = \frac{9 \kappa^2}{16\lam^2} \sqrt{|\nu|B_\nu}\left( |\nu + 2\lam| \sqrt{|\nu -2\lam| B_{\nu -2\lam}}  + |\nu - 2\lam| \sqrt{|\nu + 2\lam|B_{\nu + 2\lam}} \right).
\label{24}
\en
As in the MMO model, $K_{\nu, 0} = K_{0, \nu} = 0$. 

For $|\nu| \gg \lam$, all these models reduce to the Wheeler-DeWitt equation \eqref{7}.   

\section{Bohmian loop quantum cosmology}\label{bohmian lqg}
In the Bohmian theory there is again an actual scalar field and an actual metric of the form \eqref{1}. Since the gravitational part of the wave equation \eqref{11} is now a difference operator, rather than a differential operator, we need to  develop a Bohmian theory which results in a jump process rather than a deterministic process. Such processes have been introduced in the context of quantum field theory to account for particle creation and annihilation \cite{bell84,duerr032,duerr05a}. In the Bohmian theory, the scalar field will evolve continuously, while the scale factor $a$, which will be expressed in terms of $\nu$ using \eqref{9}, takes discrete values, determined by $\nu = 4 n\lambda$ with $n \in {\mathbb Z}$. 

The wave equation \eqref{11} implies the continuity equation
\be
\pa_\phi J_{\nu}(\phi)  = \sum_{\nu'} J_{\nu,\nu'}(\phi) ,
\en
where
\be
J_{\nu}(\phi) = B_\nu \pa_\phi S_\nu(\phi) ,\qquad   J_{\nu, \nu'}(\phi) = -K_{\nu,\nu'} \textrm{Im}\left( \psi_\nu (\phi)\psi^*_{\nu'} (\phi)\right).
\en
$J_{\nu, \nu'}$ is anti-symmetric and non-zero only for $\nu'=\nu \pm 4\lam$ for the LQC models considered above. Writing
\be
\sum_{\nu'} J_{\nu,\nu'} = \sum_{\nu'} \left( {\widetilde T}_{\nu, \nu'} |\psi_{\nu'}|^2 - {\widetilde T}_{\nu',\nu} |\psi_{\nu}|^2\right),
\en
where
\begin{equation}
{\widetilde T}_{\nu, \nu'}(\phi) = \left\{ 
\begin{array}{ll}
\frac{J_{\nu, \nu'}(\phi)}{|\psi_{\nu'}(\phi)|^2} & \quad \text{if } J_{\nu, \nu'} (\phi)> 0\\
0 &\quad  \text{otherwise}
\end{array} \right.,
\end{equation}
we can introduce the following Bohmian dynamics which preserves the quantum equilibrium distribution $|\psi_{\nu}(\phi)|^2d\phi$. The scalar field satisfies the guidance equation
\be
\dot \phi = NC B_\nu \pa_\phi S_\nu ,
\label{40}
\en
where $\psi_\nu = |\psi_\nu|\ee^{\ii S_\nu }$. The variable $\nu$, which determines the scale factor, may jump $\nu' \to \nu$ with transition rates given by $T_{\nu, \nu'}(\phi)= NC {\widetilde T}_{\nu, \nu'}(\phi)$. That is, $T_{\nu, \nu'}(\phi)$ is the probability to have a jump $\nu' \to \nu$ in the time interval $(t,t+dt)$. Note that the jump rates at a certain time depend on both the wave function and on the value of $\phi$ at that time. The properties of $J_{\nu, \nu'}$ imply that for a fixed $\nu$ either $T_{\nu, \nu +4\lam}$ or $T_{\nu, \nu -4\lam}$ may be non-zero (not both). The jump rates are ``minimal'', i.e., they correspond to the least frequent jump rates that preserve the quantum equilibrium distribution \cite{duerr05a}. Just as in the classical case and the Bohmian Wheeler-DeWitt theory, the lapse function is arbitrary, which guarantees time-reparameterization invariance.

Since the evolution of the scale factor is no longer deterministic like in the Wheeler-DeWitt theory, but stochastic, the metric is no longer Lorentzian. Namely, once there is a jump, the metric becomes discontinuous. The metric is only ``piece-wise''  Lorentzian, i.e., Lorentzian in between two jumps.

For $|\nu| \gg \lam$ (taking the limit $\lam \to 0$), this Bohmian theory reduces to the one of the Wheeler-DeWitt equation (using similar arguments as in \cite{vink93}). That is why we have chosen the particular form \eqref{40} for the guidance equation for $\phi$. 

Let us now turn to the question of singularities. If $T_{0, \pm 4\lambda} = 0$, then the scale factor $a$ (or $\nu$) can never jump to zero, so a big crunch is not possible. If $T_{\pm 4\lambda, 0} = 0$, then the scale factor can not jump from zero to a non-zero value, so a big bang is not possible. Hence there are no singularities if $J_{0, \pm 4\lambda} = 0$. If the boundary condition $\psi_0=0$ is imposed, then $J_{0, \pm 4\lambda} = 0$ for all the LQC models that are considered here. However, except for sLQC, $J_{0, \pm 4\lambda} = 0$ without imposing this boundary condition. To see this, consider the wave equation evaluated for $\nu=0$. Since $B_0=0$ in the APS, MMO and sMMO models, this results in
\be
K_{0,4\lam} \psi_{4\lam} + K_{0,-4\lam} \psi_{-4\lam} +  K_{0,0} \psi_{0} = 0.
\en
Using the properties $K_{0,\nu} = K_{0,-\nu}$ and $\psi_\nu = \psi_{-\nu}$, we obtain that
\be
\textrm{Im}\left( \psi^*_0 K_{0,\pm 4\lam} \psi_{\pm 4\lam} \right) = 0
\en
and hence that $J_{0, \pm 4\lambda} = 0$ (even if $\psi_0 \neq 0$). This argument can not be used in sLQC since $B_0 \neq 0$ in that case. In any case, the sLQC model is considered only as a simplification of the APS model; it does not follow from applying the loop quantization techniques to mini-superspace. In summary, Bohmian loop quantum cosmology models for which the wave equation \eqref{11} has the properties that $B_0=0$, $K_{0,\nu} = K_{0,-\nu}$ and $\psi_\nu = \psi_{-\nu}$, do not have singularities. Importantly, no boundary conditions need to be assumed.  

In the case that $\psi$ is real, both $\phi$ and $a$ are static. For other possible solutions, the wave equation needs to be solved first. This is rather hard, but can perhaps be done in sLQC since the eigenstates of the gravitational part of the Hamiltonian are known in this case. Something can be said about the asymptotic behaviour however. Since for large $\nu$ this Bohmian theory reduces to the one for the Wheeler-DeWitt theory, the trajectories will tend to be classical in this regime. Namely consider solutions \eqref{5} to the Wheeler-DeWitt equation for which the functions $\psi_R$ and $\psi_L$ go to zero at infinity. Then for $\al \to \infty$, the wave functions $\psi_R$ and $\psi_L$ become approximately non-overlapping in $(\phi,\al)$-space. As such the Bohmian motion will approximately be determined by either $\psi_R$ or $\psi_L$ and hence classical motion is obtained. This implies an expanding or contracting (or static) universe. We expect that a bouncing universe will be the generic solution. 
 
So far we assumed $k=\Lambda=0$. In the case $k=\pm 1$ or $\Lambda \neq 0$ there is an extra potential term $W_\nu$ in \eqref{11}, which amounts to the replacement 
\be
K_{\nu,\nu'} \to K_{\nu,\nu'} + W_\nu \delta_{\nu,\nu'}
\en
compared to the free case. $\Lambda > 0$ is discussed in \cite{ashtekar06c,pawlowski12}, $\Lambda < 0$ in \cite{ashtekar06c}, $k=1$ in \cite{ashtekar07,szulc07a}, $k=-1$ in \cite{vandersloot07,szulc07b}. In each case, the extra term is merely a potential term so that it does not contribute to the Bohmian jump process. So the same results hold concerning singularities as in the free case.

\section{On the role of time}\label{time}
In the Bohmian Wheeler-DeWitt theory and LQC there is time-reparameterization invariance, so that time is relational. An actual clock should be modeled in terms of the other variables. In the case of the Wheeler-DeWitt theory, if either $a$ or $\phi$ is monotonically increasing with time (at least for some period of time), it could play the role of a clock (for that period of time). One can then express the evolution of the other variable in terms of the clock variable. In the case of LQC, the same is true, except if the scale factor is taken as a clock variable, then it will be a discrete one.

The situation is different in the standard quantum mechanical approach to the Wheeler-DeWitt theory and loop quantum theory. There one has to deal with the notorious problem of time \cite{isham92,kuchar92,kiefer04}. In both cases, the wave equation does not depend on time and hence the wave function is static. So how does one account for apparent time evolution? One attempt for a solution is to take one of the variables, say $\phi$, as time and to take the square root of the wave equation \eqref{11}, which results in a Schr\"odinger-like equation 
\be
\ii \pa_\phi \psi^{\pm}(\phi,\nu) = \pm  \Theta \psi^{\pm}(\phi,\nu)
\label{50}
\en
in the case of the Wheeler-DeWitt theory and in
\be
\ii \pa_\phi \psi^{\pm}_{\nu}(\phi) = \pm \sum_{\nu'} \Theta_{\nu,\nu'} \psi^{\pm}_{\nu'}(\phi)
\label{51}
\en
in the case of LQC \cite{ashtekar06b,ashtekar06c,ashtekar06d,ashtekar08}. However, also the scale factor could be taken as clock variable. In particular, in the case of the Wheeler-DeWitt theory without potentials, the wave equation is completely symmetric with respect to interchange of $\al$ and $\phi$, and therefore there seems to be no reason to prefer either one as a time variable. The argument to take $\phi$ as time variable in \cite{ashtekar06b,ashtekar06c,ashtekar06d,ashtekar08} is that {\em classically} it is monotonic (just as the scale factor). However, we believe that the classical behaviour of some variables should have no implications concerning the suitability to act as time variables in the quantum case. Different time variables also lead to different theories (characterized by different Hilbert spaces) and in particular to different conclusions concerning the presence of singularities.

In the Bohmian treatments no such ambiguities arise. In particular, whether variables act as clock variables depends on their quantum behaviour, not on their classical behaviour. The wave equations \eqref{50} and \eqref{51} (or similar ones) could in principle be derived as effective equations (using the conditional wave function \cite{duerr92a,duerr09}), when the Bohmian variable $\phi$ behaves as a clock variable. 

So on the fundamental level, the scalar field is not regarded as a time variable in the Bohmian theory. However, for the sake of comparison to the analysis of singularities in the Wheeler-DeWitt theory in the context of standard quantum mechanics \cite{ashtekar06b,ashtekar06c,ashtekar06d,ashtekar08} and consistent histories \cite{craig10,craig11,craig13b,craig16}, for which the starting point is \eqref{50}, an alternative Bohmian model was considered \cite{falciano15} where the variable $\phi$ is also regarded as a time variable from the start. In this model, the equilibrium measure $|\psi(\phi,\al)|^2d\al$ is normalizable which hence directly allows for probabilistic statements. It was found that the probability $P_s$ for a trajectory $\al(\phi)$ to develop a singularity satisfies $ 1/2 \leqslant P_s \leqslant 1$. So, just as in the fundamental Bohmian theory, singularities may or may not exist depending on the wave function and the initial conditions.

One can do a similar analysis for LQC. Let us first consider the Hilbert space. The Hilbert space is given by ${\mathcal H}={\mathcal H}^{+} \oplus {\mathcal H}^{-}$, with ${\mathcal H}^{+}= {\mathcal H}^{-}$ the positive and negative frequency Hilbert spaces. The inner product on ${\mathcal H}^{\pm}$ is given by $\langle \psi | \chi \rangle = \sum_{\nu \in 4\lam {\mathbb{Z}}} \psi^*_\nu B_\nu \chi_\nu$. (The operator $\Theta$ needs to be properly defined in order to be self-adjoint with respect to this inner product \cite{kaminski09}.) By taking this inner product, states can not have support on $\nu=0$. This is because if $B_0 = \infty$, like in sLQC then, states $\psi_0$ have infinite norm and if $B_0 = 0$, like in the other models, then the states $\psi_0$ have zero norm. So for two states $\Psi = (\psi^+,\psi^-)$ and $X = (\chi^+,\chi^-)$ the inner product in ${\mathcal H}$ is 
\be
\langle \Psi | X \rangle = \sum_{\nu \in 4\lam {\mathbb{Z}}} \left[ (\psi^+_\nu)^* B_\nu \chi^+_\nu + (\psi^-_\nu)^* B_\nu \chi^-_\nu \right] .
\en
Since states have no support on $\nu=0$, singularities are immediately eliminated in the context of standard quantum mechanics. (This point is not really emphasized in \cite{ashtekar06b,ashtekar06c,ashtekar06d,ashtekar08}. There, the issue of singularities is analyzed by considering whether the wave function is peaked around the singularity \cite{ashtekar06b,ashtekar06c,ashtekar06d} or whether the expectation value of the volume operator becomes zero \cite{ashtekar08}.) Also in the Bohmian theory the singularities are eliminated. Because of the choice of inner product it is natural to take the quantum equilibrium distribution to be $P_\nu(\phi) = B_\nu (|\psi^+_\nu(\phi)|^2 + |\psi^-_\nu(\phi)|^2) $, which is now normalized to one. This distribution satisfies the continuity equation
\be
\pa_\phi P_\nu  = \sum_{\nu'} J_{\nu,\nu'},
\en
where now
\be
J_{\nu,\nu'} = 2 B_\nu  {\textrm{Im}} \left[(\psi^+_\nu)^* \Theta_{\nu,\nu'} (\psi^+_{\nu'})- (\psi^-_\nu)^* \Theta_{\nu,\nu'} \psi^-_{\nu'}\right] .
\en
The jump rates to have a jump $\nu' \to \nu$ in the time interval $(\phi,\phi+d\phi)$ are given by
\begin{equation}
T_{\nu, \nu'}(\phi) = \left\{ 
\begin{array}{ll}
\frac{J_{\nu, \nu'}(\phi)}{P_{\nu'}(\phi)} & \quad \text{if } J_{\nu, \nu'} (\phi)> 0\\
0 &\quad  \text{otherwise}
\end{array} \right..
\end{equation}
For large $\nu$ ($\lam \to 0$), this Bohmian model reduces to the one introduced in \cite{falciano15} for the square root of the Wheeler-DeWitt equation, at least for the square-root expressions given by Ashtekar {\em et al.}\ \cite{ashtekar06d,ashtekar08}. In \cite{martin-benito09,banerjee12}, Mart\'in-Benito {\em et al.}\ consider a different square-root expression, which leads to a different square root of the Wheeler-DeWitt equation.

There will be no big bang or big crunch singularity if $T_{0, \nu} = T_{\nu, 0} =0$ and this is trivially guaranteed since $\psi^+_0=\psi^-_0=0$.

\section{Modified Wheeler-DeWitt equation}\label{modified}
In section \ref{lqg}, we found that Bohmian loop quantum cosmology models for which the wave equation \eqref{11} has the properties that $B_0=0$, $K_{0,\nu} = K_{0,-\nu}$ and $\psi_\nu = \psi_{-\nu}$, do not have singularities. Consider now a modified Wheeler-DeWitt equation
\be
B(\nu) \pa^2_\phi \psi - 9\kappa^2 \pa_\nu (|\nu| \pa_\nu \psi) = 0, 
\label{70}
\en
where $\psi(\phi,-\nu)=\psi(\phi,\nu)$ and where $B(\nu)$ is some function that approximates $1/|\nu|$ in the limit of large $|\nu|$ and satisfies $B(0)=0$. The guidance equations in this case are
\be
\dot \phi = NC B(\nu) \pa_\phi S , \quad \dot \nu = - N9C \kappa^2 |\nu| \pa_\nu S .
\label{71}
\en
Is such a modification sufficient to eliminate the singularities? Or is the discreteness also essential? Or does it depend on the choice of the function $B$?

As a simple example we can consider a function $B$ which is zero for $|\nu| < \nu_0 $ for some $\nu_0>0$, which we can take arbitrarily small. Then for $|\nu| \leqslant \nu_0$, the wave equation reduces to $\pa_\nu (|\nu| \pa_\nu \psi) = 0$, which implies $\psi(\phi,\nu) = \chi_1(\phi) {\textrm{sgn}}(\nu) \ln(|\nu|) + \chi_2(\phi)$, with the $\chi_i$ some functions of $\phi$. Since $\psi(\phi,-\nu)=\psi(\phi,\nu)$, it must be that $\chi_1=0$ and therefore the guidance equation for $\nu$ implies that $\nu$ must be static. Hence in this case there are no singularities.

Note that the condition $\psi(\phi,-\nu)=\psi(\phi,\nu)$ actually entails the boundary condition $\pa_\nu\psi(\phi,0)=0$. If we consider the equations \eqref{70} and \eqref{71} just for $\nu \geqslant 0$, which would be natural, then without the condition $\pa_\nu\psi(\phi,0)=0$ singularities are still possible. Namely, for $\nu < \nu_0 $, \eqref{71} implies that $\phi$ is constant, while ${\dot \nu}= -N9C \kappa^2 {\textrm{Im}}(\chi^*_2(\phi) \chi_1(\phi))/|\chi_1(\phi) \ln(\nu) + \chi_2(\phi)|^2$. Taking for example $\chi_1(\phi)=-i$ and $\chi_2(\phi)=1$ for the constant value of $\phi$ that is obtained, integration of guidance equation for $\nu$ yields $9C \kappa^2 \tau=\nu(3+\ln \nu(\ln \nu - 2)) + 9C\kappa^2c $, with $\tau$ the proper time for a co-moving observer (see section \ref{wdwbm}) and $c$ an integration constant. But this implies that the singularity is reached (i.e.\ $\nu=0$) for $\tau = c$.

In summary, singularities can be eliminated by assuming the above mentioned modifications to the Wheeler-DeWitt equation. It is unclear whether this is so for any choice of $B$ which satisfies the above mentioned assumptions.

Finally, from the guidance equations \eqref{71} and the modified Wheeler-DeWitt equation \eqref{70} the modified Friedmann equation follows
\be
\frac{\dot a^2}{N^2a^2} = 2\ka^2 \left( \frac{\dot \phi^2}{2N^2 |\nu B|} + Q_M \right) +2 Q_G ,
\en
where
\be
Q_M = - \frac{C^2 B}{2|\nu||\psi|} \pa^2_\phi|\psi| , \qquad Q_G = \frac{9C^2\ka^4}{2|\nu||\psi|} \pa_\nu (|\nu|\pa_\nu |\psi|)  
\en
are respectively the matter and graviational quantum potential. The presence of the quantum potentials allows $\dot a$ to become zero and hence for the universe to undergo a bounce. Regarding the present model as a continuum approximation of LQG one can perhaps deduce an effective equation that illustrates a generic bouncing behaviour, in analogy with the one considered in \cite{ashtekar11}.

\section{Conclusion}
We have developed a Bohmian version for some common models of LQC for homogeneous and isotropic FLRW space-time coupled to a homogeneous scalar field, with arbitrary values of the constant spatial curvature and cosmological constant. The Bohmian theory describes an actual metric, whose sole degree of freedom in this case is the scale factor, and a scalar field, whose dynamics depends on the wave function. While the scalar field evolves continuously in time, the scale factor changes stochastically (unlike in the case of the Wheeler-DeWitt theory where it changes deterministically). We showed that a non-zero scale factor never jumps to zero and conversely that a zero scale-factor never becomes non-zero. This means that there is no big bang or big crunch singularity. This result was obtained without assuming any boundary conditions; it follows solely from the structure of the wave equation. (Sometimes boundary conditions are considered to avoid singularities, like for example the condition that the wave function vanishes on singular metrics \cite{dewitt67a,kiefer04}.) 

It is to be expected that similar results hold for other space-times, such as anisotropic space-times, like for example the Kantowski-Sachs space-time, which is particularly interesting since it can be used to describe the interior of a black hole and hence can be used to study the question of black-hole singularities \cite{ashtekar06a}.

So far we have restricted our attention merely to the question of singularities. While we have established that there are no singularities, further work is required to learn the generic behaviour of solutions.

\section{Acknowledgments}
This work was supported by the Deutsche Forschungsgemeinschaft. It is a pleasure to thank Alejandro Corichi, Felipe Falciano, Claus Kiefer, Hannah Ochner, Nelson Pinto-Neto, Parampreet Singh, and especially Detlef D\"urr, for useful discussions and comments. 

\section{Additional Information}
Competing Interests: The author declares no competing financial interests.

\end{document}